\documentclass[12pt]{amsart}
\usepackage{amsmath,amssymb,amsbsy,amsfonts,latexsym,amsopn,amstext,
                                               amsxtra,euscript,amscd}

\newfont{\teneufm}{eufm10}
\newfont{\seveneufm}{eufm7}
\newfont{\fiveeufm}{eufm5}
\newfam\eufmfam
                \textfont\eufmfam=\teneufm \scriptfont\eufmfam=\seveneufm
                \scriptscriptfont\eufmfam=\fiveeufm
\def\bbbc{{\mathchoice {\setbox0=\hbox{$\displaystyle\rm C$}\hbox{\hbox
to0pt{\kern0.4\wd0\vrule height0.9\ht0\hss}\box0}}
{\setbox0=\hbox{$\textstyle\rm C$}\hbox{\hbox
to0pt{\kern0.4\wd0\vrule height0.9\ht0\hss}\box0}}
{\setbox0=\hbox{$\scriptstyle\rm C$}\hbox{\hbox
to0pt{\kern0.4\wd0\vrule height0.9\ht0\hss}\box0}}
{\setbox0=\hbox{$\scriptscriptstyle\rm C$}\hbox{\hbox
to0pt{\kern0.4\wd0\vrule height0.9\ht0\hss}\box0}}}}
\def\bbbq{{\mathchoice {\setbox0=\hbox{$\displaystyle\rm
Q$}\hbox{\raise 0.15\ht0\hbox to0pt{\kern0.4\wd0\vrule
height0.8\ht0\hss}\box0}} {\setbox0=\hbox{$\textstyle\rm
Q$}\hbox{\raise 0.15\ht0\hbox to0pt{\kern0.4\wd0\vrule
height0.8\ht0\hss}\box0}} {\setbox0=\hbox{$\scriptstyle\rm
Q$}\hbox{\raise 0.15\ht0\hbox to0pt{\kern0.4\wd0\vrule
height0.7\ht0\hss}\box0}} {\setbox0=\hbox{$\scriptscriptstyle\rm
Q$}\hbox{\raise 0.15\ht0\hbox to0pt{\kern0.4\wd0\vrule
height0.7\ht0\hss}\box0}}}}
\def\bbbt{{\mathchoice {\setbox0=\hbox{$\displaystyle\rm
T$}\hbox{\hbox to0pt{\kern0.3\wd0\vrule height0.9\ht0\hss}\box0}}
{\setbox0=\hbox{$\textstyle\rm T$}\hbox{\hbox
to0pt{\kern0.3\wd0\vrule height0.9\ht0\hss}\box0}}
{\setbox0=\hbox{$\scriptstyle\rm T$}\hbox{\hbox
to0pt{\kern0.3\wd0\vrule height0.9\ht0\hss}\box0}}
{\setbox0=\hbox{$\scriptscriptstyle\rm T$}\hbox{\hbox
to0pt{\kern0.3\wd0\vrule height0.9\ht0\hss}\box0}}}}
\def\bbbs{{\mathchoice
{\setbox0=\hbox{$\displaystyle     \rm S$}\hbox{\raise0.5\ht0\hbox
to0pt{\kern0.35\wd0\vrule height0.45\ht0\hss}\hbox
to0pt{\kern0.55\wd0\vrule height0.5\ht0\hss}\box0}}
{\setbox0=\hbox{$\textstyle        \rm S$}\hbox{\raise0.5\ht0\hbox
to0pt{\kern0.35\wd0\vrule height0.45\ht0\hss}\hbox
to0pt{\kern0.55\wd0\vrule height0.5\ht0\hss}\box0}}
{\setbox0=\hbox{$\scriptstyle      \rm S$}\hbox{\raise0.5\ht0\hbox
to0pt{\kern0.35\wd0\vrule height0.45\ht0\hss}\raise0.05\ht0\hbox
to0pt{\kern0.5\wd0\vrule height0.45\ht0\hss}\box0}}
{\setbox0=\hbox{$\scriptscriptstyle\rm S$}\hbox{\raise0.5\ht0\hbox
to0pt{\kern0.4\wd0\vrule height0.45\ht0\hss}\raise0.05\ht0\hbox
to0pt{\kern0.55\wd0\vrule height0.45\ht0\hss}\box0}}}}
\def\bbbz{{\mathchoice {\hbox{$\sf\textstyle Z\kern-0.4em Z$}}
{\hbox{$\sf\textstyle Z\kern-0.4em Z$}} {\hbox{$\sf\scriptstyle
Z\kern-0.3em Z$}} {\hbox{$\sf\scriptscriptstyle Z\kern-0.2em
Z$}}}}

\newtheorem{theorem}{Theorem}
\newtheorem{lemma}[theorem]{Lemma}

\newtheorem{cor}[theorem]{Corollary}

\def\squareforqed{\hbox{\rlap{$\sqcap$}$\sqcup$}}
\def\qed{\ifmmode\squareforqed\else{\unskip\nobreak\hfil
\penalty50\hskip1em\null\nobreak\hfil\squareforqed
\parfillskip=0pt\finalhyphendemerits=0\endgraf}\fi}

\def\cA{{\mathcal A}}
\def\cB{{\mathcal B}}

\def\cE{{\mathcal E}}

\def\cP{{\mathcal P}}

\def\cR{{\mathcal R}}

\def\cW{{\mathcal W}}

\def \sf {\mathfrak s}

\newcommand{\ignore}[1]{}

\def\mand{\qquad\mbox{and}\qquad}

\def\\{\cr}
\def\({\left(}
\def\){\right)}

\def\be{\begin{equation}}
\def\ee{\end{equation}}

\def\vp{\varphi}
\def\arrowk{^\to{\kern -6pt\topsmash k}}
\def\arrowK{^{^\to}{\kern -9pt\topsmash K}}
\def\arrowt{^\to{\kern -6pt\topsmash t}}
\def\arrowr{^\to{\kern-6pt\topsmash r}}
\def\arrowvp{^\to{\kern -8pt\topsmash\vp}}
\def\tk{\tilde{\kern 1 pt\topsmash k}}
\def\barm{\bar{\kern-.2pt\bar m}}
\def\barN{\bar{\kern-1pt\bar N}}
\def\barA{\, \bar{\kern-3pt \bar A}}

\def\be{\begin{equation}}
\def\ee{\end{equation}}
\begin{document}




\title[Evasive Properties of Sparse Graphs and Primes]{Evasive Properties of Sparse Graphs
and Some Linear Equations in Primes}
%
\author[I. E. Shparlinski]
{Igor E. Shparlinski}
\address{Department of Computing, Macquarie University, Sydney, NSW 2109, Australia}
\email{igor.shparlinski@mq.edu.au}


\begin{abstract} We give an unconditional version of a conditional, 
on the Extended Riemann Hypothesis, result of 
L.~Babai, A.~Banerjee, R.~Kulkarni and V.~Naik (2010) on the evasiveness 
of sparse graphs. 
\end{abstract}
\maketitle

\section{Introduction}

A Boolean function of $m$ variables is called {\it evasive\/}
if its deterministic query (decision-tree) complexity is $m$. 

A graph property $\cP_n$ of $n$-vertex graphs is a collection of graphs on the vertex set
$\{1, \ldots, n\}$ 
that is invariant under relabelling of the vertices. A property $\cP_n$ is called
monotone if it is preserved under the deletion of edges. The trivial graph
properties are the empty set and the set of all graphs. 
We say that $\cP_n$ is {\it evasive\/} if the Boolean function on 
$$
m = \frac{n(n-1)}{2}
$$
variables, deciding whether an  $n$-vertex graph given by the 
adjacency matrix belongs to $\cP_n$, is evasive. 

The famous {\it Karp Conjecture} asserts that 
any monotone nontrivial graph property is evasive, see~\cite{BBKN,
ScTr} and references therein. 

Towards this conjecture, Babai,  Banerjee, Kulkarni and Naik~\cite[Theorem~1.4~(b)]{BBKN} have shown that, 
under the Extended Riemann Hypothesis,   
for any fixed $\varepsilon> 0$, any 
nontrivial monotone property of graphs on  $n$ vertices  with at most
$n^{5/4- \varepsilon}$ edges is evasive for a sufficiently large $n$.

The unconditional result of~\cite[Theorem~1.4~(c)]{BBKN} is much weaker, 
and applies to graphs with 
at most $cn\log n$ edges (for some absolute constant $c > 0$). 

Furthermore, under the so-called {\it Chowla Conjecture\/} about the 
smallest prime in an arithmetic progression (which goes far beyond
of what the Extended Riemann Hypothesis immediately implies), 
Babai,  Banerjee, Kulkarni and Naik~\cite[Theorem~1.4~(a)]{BBKN} show that 
for any fixed $\varepsilon> 0$, any 
nontrivial monotone property of graphs on  $n$ vertices  with at most
$n^{3/2- \varepsilon}$ edges is evasive for a sufficiently large $n$.

These estimates rely on some results about the distribution of 
primes in arithmetic progressions. 
 Here we show that the 
{\it Bombieri-Vinogradov theorem\/}, see~\cite[Theorem~17.1]{IwKow}, is sufficient to replace  the Extended Riemann Hypothesis
and so we obtain the same result unconditionally which 
improves~\cite[Theorem~1.4~(c)]{BBKN} that gives the evasiveness for 
graphs on  $n$ vertices  with at most $n\log n$ edges. 

\begin{theorem}
\label{thm:Evasive-1} There is a 
function $f(n) =  n^{5/4 + o(1)}$ such that
any nontrivial monotone property of graphs on $n$ vertices, with at most
$f(n)$ edges, is evasive for a sufficiently large $n$.
\end{theorem}

Furthermore, we show that using a different approach, based on a  result
of Balog and S{\'a}rk{\"o}zy~\cite{BaSa}
about prime divisors of sum-sets (which in turn is based on sieve
methods), one can obtain much stronger estimates that hold for almost 
all $n$.   


We need to introduce some notation. 
For an integer $k$, we use $P(k)$ to denote the largest prime divisor of
$k$ (we also set $P(1)=0$).

\begin{theorem}
\label{thm:Evasive-2} Assume that for some real positive $\alpha < 1$ and $A$ we have
$$
\# \{r \le z~:~ r~\text{prime}, \ P(r-1) > r^\alpha\} \ge A \frac{z}{\log z}
$$
as $z\to \infty$. 
Then for any positive $\gamma \le \alpha$ 
there is a constant $c(\alpha,\gamma, A) > 0$ that depends only on $\alpha$, 
$\gamma$ and $A$ such that  for all but at most
$O\(x^{\max\{0, 2\gamma - 1\}} (\log x)^{4}\)$ integers $n\le x$
any nontrivial monotone property of graphs on $n$ vertices with at most
$c(\alpha,\gamma, A)n^{1+\gamma}$ edges, is evasive. 
\end{theorem}

The standard heuristic suggests that the condition of Theorem~\ref{thm:Evasive-2}
holds with any $\alpha< 1$. 
Unconditionally, by a result of Baker and Harman~\cite{BaHa}, it is known that we can
take
$$ 
\alpha =0.677
$$
for some $A> 0$.  
Thus, with $\gamma = \alpha = 0.677$ we derive:

\begin{cor}
\label{cor:Evasive-1} 
There is an absolute constant $c > 0$, such that for all but at most 
$O\(x^{0.354}(\log x)^{4}\)$ integers $n\le x$
any nontrivial monotone property of graphs on $n$ vertices, with at most
$c n^{1.677}$ edges, is evasive.
\end{cor} 
 
Finally, taking $\gamma = 1/2$ and $\alpha = 0.677$ in  
Theorem~\ref{thm:Evasive-2}, we obtain
an unconditional version of the bound of~\cite[Theorem~1.4~(a)]{BBKN} 
hiowever with a small exceptional set. 

\begin{cor}
\label{cor:Evasive-2} 
There is an absolute constant $c > 0$, such that for all but at most 
$O\((\log x)^{4}\)$ integers $n\le x$
any nontrivial monotone property of graphs on $n$ vertices, with at most
$c n^{3/2}$ edges, is evasive.
\end{cor}

We note that in~\cite[Theorem~1.4~(a)]{BBKN} 
the bound of Corollary~\ref{cor:Evasive-2} (in a slightly weaker form 
$n^{3/2 -\varepsilon}$ for any $\varepsilon>0$) 
is established for all sufficiently large $n$ under the so-called 
{\it Chowla Conjecture\/}. However proving this conjecture seems to be far beyond 
the capabilities of the modern number theory.

\section{Preparations}

Throughout the paper, the implied constants in the symbols `$O$',  `$\ll$' 
and `$\gg$'  may occasionally,
where obvious, depend  on the small real parameter $\varepsilon>0$
and are absolute otherwise. 
We recall that the notations $U = O(V)$,  $U \ll V$  and $V \gg U$ are all
equivalent to the assertion that the inequality $|U|\le c|V|$ holds for some
constant $c>0$.

Our main technical tool is the following result obtained 
and used in~\cite[Section~5]{BBKN}. 
For an integer $n\ge 1$ we define the function 
\begin{equation}
\label{eq:fn}
f(n) = \max_{(k,p,q,r)\in \cW_n}\min\left\{p^2k, pkr, qr\right\}, 
\end{equation}
where the maximun is taken over the set $\cW_n$ of all quadruples $(k,p,q,r)$ of 
integers $k \ge 1$ and primes $p$, $q$, $r$ with 
\begin{equation}
\label{eq:cond}
n = kp + r \mand r \equiv 1 \pmod q.
\end{equation}

\begin{lemma}
\label{lem:BBKN} 
There is an absolute constant $c>0$ such that   any 
nontrivial monotone properties of graphs on  $n$ vertices  with at most
$cf(n)$ edges is evasive for a sufficiently large $n$.
\end{lemma}

In~\cite[Section~5]{BBKN} individual results about the distribution 
of primes in arithmetic progressions, have been obtained to get lower bounds on 
$f(n)$ and thus on the evasiveness of sparse graphs. 

Here we use several  results about the distribution of 
primes in arithmetic on average to improve the estimates from~\cite[Section~5]{BBKN}. 

For integers $m > a \ge 0$ and a real $y> 0$, let 
$$
\psi(y;m,a) = \sum_{n \le y}\Lambda(y), 
$$
where, as usual, $\Lambda$ denotes the von~Mangoldt function given by
$$
\Lambda(n)=
\begin{cases}
\log p&\quad\text{if $n$ is a power of the prime $p$,} \\
0&\quad\text{if $n$ is not a prime power.}
\end{cases}
$$
We also use $\varphi(m)$ to denote the Euler function of $m$. 

We now recall (a somewhat simplified) version of the Bombieri-Vinogradov 
theorem, see~\cite[Theorem~17.1]{IwKow}.

\begin{lemma}
\label{lem:B-V} 
For every $A> 0$ there exists $B$ such that 
for any real $z > 1$, 
$$
\sum_{m \le  \sqrt{z}/(\log z)^B}
\max_{y \le z} \max_{\gcd(a,m)=1}\left|\psi(y;m,a)-\frac{ y}{\varphi(m)}\right|  \ll \frac{z}{(\log z)^A}.
$$
\end{lemma}

Finally, by a straight-forward modification of a result of 
Balog and S{\'a}rk{\"o}zy~\cite[Theorem~2]{BaSa} (which in the 
original formulation applies to $P(a+b)$ rather than to $P(a-b)$)
we have: 

\begin{lemma}
\label{lem:BS} There is an absolute constant  $c>0$ such that for any sets $\cA, \cB \subseteq \{1, \ldots, N\}$ with 
$$
\# \cA \# \cB \ge c  N (\log N)^{2}
$$
we have 
$$
\max_{a\in \cA, \, b \in \cB} P(a-b) 
\gg \frac{(\# \cA \# \cB )^{1/2}}{\log N }.
$$
\end{lemma}

We recall that when both sets $\cA$ and $\cB$ are large (of cardinalities 
of order $N$) an improvement of 
Lemma~\ref{lem:BS} is given by S{\'a}rk{\"o}zy and  Stewart~\cite{SaSt},
see also a survey of releted results given by Stewart~\cite{Stew}.

\section{Proof of Theorem~\ref{thm:Evasive-1}}

Let us fix some $\varepsilon> 0$, and consider the products 
$m = pq$ where $p$ and $q$ are distinct primes from the interval
$[n^{1/4-\varepsilon}, 2n^{1/4-\varepsilon}]$. 
Clearly for some constant $c > 0$ there are at least $M_1 \ge c n^{1/2-2\varepsilon}/(\log x)^2$
such values of $m$. 
On the other hand, by Lemma~\ref{lem:B-V}, 
applied with $A=3$, 
 we see that the number $M_2$ of $m \in [n^{1/2-2\varepsilon}, 4n^{1/2-2\varepsilon}]$
with 
$$
\max_{y \le n/2} \max_{\gcd(a,m)=1}\left|\psi(y;m,a)-\frac{y}{\varphi(m)}\right| \ge \frac{n}{10m}
$$
satisfies
$$
M_2 \frac{n}{4n^{1/2-2\varepsilon}}
\ll \frac{n}{(\log n)^3}, 
$$
or
$$
M_2 \ll \frac{4n^{1/2-2\varepsilon}}{(\log n)^3}.
$$
Hence $M_2 < M_1$ for a sufficiently large $n$.
We now choose any two distinct primes 
$p,q \in [n^{1/4-\varepsilon}, 2n^{1/4-\varepsilon}]$
such that for $m = pq$ we have 
$$
\max_{y \le n/2} \max_{\gcd(a,m)=1}\left|\psi(y;m,a)-\frac{y}{\varphi(m)}\right| < \frac{n}{10m}.
$$
In particular, if for these $p$ and $q$ we define $a\in [0, m-1]$
by the congruences 
$$a \equiv n \pmod p \mand 
a \equiv 1 \pmod q
$$
we have 
$$
\psi(n/2;m,a) - \psi(n/4;m,a)\ge \frac{n}{4\varphi(m)}-   \frac{n}{5m} > 0.
$$
Thus there is a prime $r \in [n/4, n/2]$ with $r \equiv a \pmod {pq}$.
Setting $k = (n-r)/p$,  we obtain a representation of the form~\eqref{eq:cond}
which implies that for the function~\eqref{eq:fn}
$$
f(n) \gg n^{3/4-\varepsilon}.
$$
Since $\varepsilon$ is arbitrary, by Lemma~\ref{lem:BBKN}, 
the result now follows.

\section{Proof of Theorem~\ref{thm:Evasive-2}}
Clearly it is enough to show that the result holds for
all but  possibly $O\(x^{\max\{0, 2\gamma - 1\}} (\log x)^{3}\)$ 
integers $n\in [x/2,x]$.

From the definition of $\alpha$ we see that there is a constant $c_0> 0$ 
such that for the set 
$$
\cR = \{r \in [c_0 x, x/4]~:~ r~\text{prime}, \ P(r-1) > r^\alpha\} 
$$
we have 
\begin{equation}
\label{eq:R}
\# \cR \gg x/\log x. 
\end{equation}

Assume that for an integer $n \in [x/2,x]$ there is with $r \in \cR$
with $P(n-r) \ge n^{\gamma}$.
Taking $p = P(n-r)$, $q = P(r-1)$ and writing $n = pk + r$, 
we see that $pk > n/2$. Thus for the function~\eqref{eq:fn} we have
$$
f(n) \gg  \min\left\{n p, nr, r^{1+\alpha}\right\} 
\gg  n^{1+\gamma}.
$$

Let $\cE$ be the set of remaining integers $n \in [x/2,x]$ for which for all  $r \in \cR$
we have $P(n-r) \le n^{\gamma}$. 
We see from Lemma~\ref{lem:BS}, applied with $\cA = \cE$ and $\cB = \cR$, that 
for any $\varepsilon> 0$ we have  
either 
$$
\# \cE \# \cR \le x(\log x)^{2}
$$
or 
$$
 \frac{(\# \cE \# \cR  )^{1/2}}{\log x} \ll x^{\gamma} .
$$
Thus, recalling~\eqref{eq:R}, we see that
$$
\#\cE\ll  x^{\max\{0, 2\gamma - 1\}} (\log x)^{3}
$$
and the result now follows.

\section{Comments}

Clearly the exponent $5/4$ in Theorem~\ref{thm:Evasive-1}
comes from the limit $z^{1/2+o(1)}$ of averaging 
in the Bombieri-Vinogradov 
theorem, see Lemma~\ref{lem:B-V}. However under the {\it Elliott-Halberstam conjecture\/},
which essentially asserts that the averaging in  Lemma~\ref{lem:B-V} 
can be extended up to $z^{1-\varepsilon}$ for any fixed $\varepsilon>0$,
see~\cite[Section~17.1]{IwKow}, allows to replace $5/4$ with $3/2$. 
This is the same result as the one obtained in~\cite{BBKN}
under the Chowla conjecture. Note that the Chowla conjecture applies 
to individual progressions and thus may be more difficult to 
establish than the Elliott-Halberstam conjecture. 
Furthermore, under the Elliott-Halberstam conjecture, 
one can take any $\alpha< 1$ in Theorem~\ref{thm:Evasive-2}

Finally, we recall that there are stronger versions of this results due
to Bombieri, Friedlander and Iwaniec~\cite{BFI1,BFI2,BFI3} 
and Mikawa~\cite{Mik}, see also a recent result of Fourvry~\cite{Fou}.
Unfortunately all these results require some restrictions
on the residues classes $a$ in $\psi(y,m,a)$ to which they apply.
This makes them difficult to use for our purpose. 

 \section*{Acknowledgements}
 
The author is very grateful to L{\'a}szl{\'o} Babai, John Friedlander and Raghav Kulkarni 
for a number of very useful discussions.

This work was initiated when the author was visiting the Centre for Quantum Technologies
at National University of Singapore and triggered by the very enthralling seminar 
talk by Raghav Kulkarni. 
The hospitality and support of this institution are gratefully acknowledged.

During the preparation of this work the author was supported in part by 
the  Australian Research Council  Grant~DP130100237.

\end{document}